\documentclass[11pt,a4paper,useAMS,usenatbib]{emulateapj}
\usepackage{epsfig}
\usepackage{amsmath}
\usepackage{natbib}
\usepackage{ulem} 
\bibpunct{(}{)}{;}{a}{}{,}

\usepackage[dvipsnames]{xcolor}

\begin{document}

\title{The Archetypal Ultra-diffuse Galaxy, Dragonfly 44, is not a Dark Milky Way}

\author{ \'Akos Bogd\'an\altaffilmark{1}}
\affil{\altaffilmark{1}Harvard-Smithsonian Center for Astrophysics, 60
Garden Street, Cambridge, MA 02138, USA; abogdan@cfa.harvard.edu}
\shorttitle{THE DARK MATTER HALO MASS OF DF~44}
\shortauthors{\'A. BOGD\'AN}

\begin{abstract}
Due to the peculiar properties of ultra-diffuse galaxies (UDGs), understanding their origin presents a major challenge. Previous X-ray studies demonstrated that the bulk of UDGs lack substantial X-ray emission, implying that they reside in low-mass  dark matter halos. This result, in concert with other observational and theoretical studies, pointed out that most UDGs belong to the class of dwarf galaxies. However, a subset of UDGs is believed to host a large population of globular clusters (GCs), which is indicative of massive dark matter halos. This, in turn, hints that some UDGs may be failed $L_{\star}$ galaxies. In this work, I present \textit{Chandra} and \textit{XMM-Newton} observations of two archetypal UDGs, Dragonfly~44  and DF~X1, and I constrain their dark matter halo mass based on the X-ray emission originating from hot gaseous emission and from the population of low-mass X-ray binaries (LMXBs) residing in GCs. Both Dragonfly~44 and DF~X1 remain undetected in X-rays. The upper limits on the X-ray emission exclude the possibility that these galaxies reside in massive ($M_{\rm vir} \gtrsim 5\times10^{11}  \ \rm{M_{\odot}}$) dark matter halos, suggesting that they are not failed $L_{\star}$ galaxies. These results demonstrate that even these iconic UDGs resemble to dwarf galaxies with $M_{\rm vir} \lesssim 10^{11}  \ \rm{M_{\odot}}$, implying that UDGs represent a single galaxy population. \\
\end{abstract}


\section{Introduction}
\label{sec:intro}

The physical characteristics of ultra-diffuse galaxies (UDGs) signify that they may represent a new class of galaxies: their surface brightness is extremely low ($ \mu_{\rm 0} (g) \gtrsim 24 \ \rm{mag \ arcsec^{-2}} $) and is similar or lower than that of dwarf galaxies, but their effective radius ($r_{\rm eff} \gtrsim 1.5$~kpc) is comparable to more massive galaxies, such as the Milky Way \citep[e.g.][]{1984AJ.....89..919S,1992AJ....103.1107S,2015ApJ...798L..45V}. In addition, stellar kinematics measurements and the discovery of an abundant population of globular clusters (GCs) around some UDGs hinted that UDGs may be the relics of massive galaxies, so-called failed $L_{\star}$ galaxies \citep{2016ApJ...828L...6V,2017ApJ...844L..11V}. However, owing to the demanding nature of deep optical follow-up measurements, only a handful of UDGs were studied in detail, which would support this exotic formation scenario. The alternative formation scenario, favored by several theoretical studies, suggested that UDGs belong to the class of dwarf galaxies, and their large spatial extent is due to energetic stellar feedback \citep[e.g.][]{2016MNRAS.459L..51A,2017MNRAS.466L...1D}. 

The crucial observational difference between these formation scenarios is the dark matter halo mass of UDGs. If UDGs are failed $L_{\star}$ galaxies, they will reside is massive dark matter halos ($M_{\rm vir} \gtrsim 5 \times 10^{11} \ \rm{M_{\odot}}$). If, however, they are genuine dwarf galaxies, they will live in a dwarf-size dark matter halos ($M_{\rm vir} \lesssim 10^{11} \ \rm{M_{\odot}}$). Recently, we explored a statistically significant sample of UDGs in isolated and in galaxy cluster environments and constrained their dark matter halo mass using X-ray observations \citep{2019ApJ...879L..12K,2020ApJ...898..164K}. These studies  demonstrate that the bulk of UDGs do not reside in massive dark matter halos, and strengthen the picture, in which UDGs are puffed-up dwarf galaxies. However, due to the statistical nature of these studies, it could not be excluded that a small subset of UDGs resides in massive dark matter halos, implying that UDGs could form via multiple channels. 

The most likely candidate UDGs with massive dark matter halos are Dragonfly~44 (hereafter DF~44) and DF~X1, which galaxies were studied in extensive follow-up campaigns. Deep optical observations suggest the existence of a substantial GC population around these galaxies, which is indicative of massive dark matter halos \citep{2016ApJ...828L...6V,2017ApJ...844L..11V,2019ApJ...880...91V}. In this picture, it is expected that UDGs will exhibit X-ray emission originating from the population of low-mass X-ray binaries (LMXBs) residing in GCs and from diffuse gaseous emission. This X-ray emission should be observable by present-day X-ray telescopes, such as \textit{Chandra} or \textit{XMM-Newton}. Both of these galaxies were subject to deep X-ray observations, which allow the detailed investigations of their X-ray emitting properties.

Recently, \citet{2020MNRAS.497.2759L} studied the ultraviolet and X-ray properties of a sample of UDGs in the Coma cluster that were discovered by the Dragonfly Telephoto Array \citep{2014PASP..126...55A}. They concluded, in agreement with \citet{2020ApJ...898..164K}, that the bulk of UDGs do not exhibit statistically significant X-ray emission. In addition, they analyzed the \textit{XMM-Newton} data of DF~44 and did not detect statistically significant unresolved emission. In this work, I present deep, high-angular resolution \textit{Chandra} observations of DF~44, the data of which can resolve LMXBs associated with GCs or a nuclear X-ray source. Based on the \textit{Chandra} data, I also constrain the X-ray luminosity originating from diffuse gaseous emission and from the population of unresolved X-ray binaries. In addition, I present the \textit{XMM-Newton} data available for DF~X1, and constrain the X-ray luminosity associated with this UDG. Overall, this study aims to fill the missing gap in our understanding about the formation scenarios of UDGs, to directly address whether the most well-studied UDGs could be failed  $L_{\star}$ galaxies, and to probe whether UDGs may have multiple formation channels. 

For the distance of DF~44 and DF~X1 I assumed $D=103$~Mpc, at which distance $1\arcsec$ corresponds to $0.476$~kpc. The Galactic absorption toward Coma cluster is $9.3\times10^{19} \ \rm{cm^{-2}}$ \citep{2016A&A...594A.116H}. Throughout this Letter I used standard $\Lambda$-CDM cosmology with   $H_0 = 71  \ \rm{km \ s^{-1} \ Mpc^{-1}},  \Omega_M=0.3$, and $\Omega_{\Lambda}=0.7$.

This Letter is structured as follows. In Section 2 I describe the analysis of \textit{Chandra} and \textit{XMM-Newton} data. I present the results in Section 3 and place these results in context in Section 4.

\begin{figure*}[ht!]
	\begin{center}
		\leavevmode
		\epsfxsize=0.92\textwidth \epsfbox{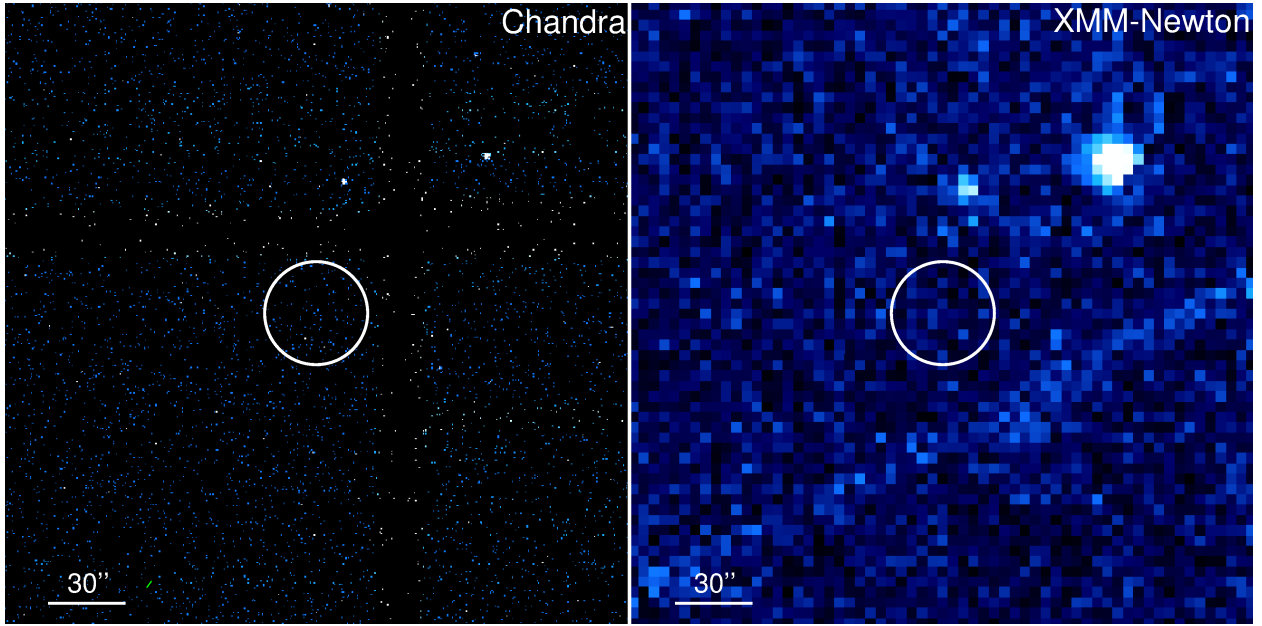}
		\vspace{0cm}
		\caption{$0.5-8$ keV band exposure corrected \textit{Chandra} (left panel) and \textit{XMM-Newton} (right panel) images of a $4\arcmin \times 4\arcmin$  region around DF~44. No diffuse X-ray emission, originating from hot gaseous emission and/or the population of unresolved X-ray binaries, is detected. In addition, no resolved X-ray sources are detected either from GC-LMXBs or from a nuclear source. The lack of statistically significant X-ray emission implies that the dark matter halo of DF~44 is not comparable with  $L_{\star}$ galaxies.} 
		\label{fig:df44}
	\end{center}
\end{figure*}

\section{Data analysis}
\label{sec:data}

\subsection{Chandra}
\label{sec:chandra}

\textit{Chandra} observed DF~44 in three ACIS-I pointings for a total exposure time of  95.4 ks. The data were analyzed using standard \textsc{CIAO}  (version 4.12) tools with \textsc{CALDB} 4.9.1.

The main steps of the analysis are similar to those outlined in \citet{2020ApJ...898..164K}. First, I reprocessed the individual observations with the \texttt{chandra\_repro} task. To filter high background time periods, I binned the light curves into 200~s bins and excluded those bins that were $2\sigma$  above the mean count rates. The original and filtered exposure times are listed in Table \ref{tab:list1}. The total clean exposure time is $76.0$~ks. 

In this work, I study the combined emission originating from hot gas and the population of LMXBs. Therefore, I carry out the analysis in the $0.5-8$~keV band (see Section \ref{sec:model} for details). To this end, I created exposure maps that reflect the spectrum of both of these components by utilizing the combination of an APEC model with $kT= 0.2$ keV temperature and metallicity of $Z=0.2 \ Z_{\rm \odot}$ \citep{2011MNRAS.418.1901B,2016ApJ...826..167G} and a power-law model with a slope of $\Gamma = 1.56$ \citep{2003ApJ...587..356I}.  Finally, I identified point sources following the procedure of \citet{2020ApJ...898..164K}. For the study of the diffuse emission, these point sources were excluded. 

To account for the background emission, I utilized a local background region around DF~44. This approach assures that not only are the instrumental and sky background components accounted for, but the large-scale emission from the Coma cluster is as well. However, this latter component does not play a major role as the emission from the intracluster emission is relatively low at a projected distance of $\sim1\degr$ (or $1.7$~Mpc) from the core of the Coma cluster.

\begin{table}
\caption{The list of analyzed \textit{Chandra} observations.}
\begin{minipage}{8.75cm}
\renewcommand{\arraystretch}{1.3}
\centering
\begin{tabular}{c c c c c c}
\hline 
Obs ID &  $T_{\rm{obs}}$ (ks) &  $T_{\rm{filt}}$ (ks) &  Instrument & Date\\
\hline
20612 & 29.7 & 22.8 & ACIS-I & 2018 Apr 5  \\
21068 & 33.1 & 29.5 & ACIS-I & 2018 Apr 6  \\
21069 & 32.6 & 23.7 & ACIS-I & 2018 Apr 8  \\ \hline
\end{tabular} 
\end{minipage}
\label{tab:list1}
\end{table}

\subsection{XMM-Newton}
\label{sec:xmm}

\begin{table}
\caption{The list of analyzed \textit{XMM-Newton} observations.}
\begin{minipage}{8.9cm}
\renewcommand{\arraystretch}{1.3}
\centering
\begin{tabular}{c c c c c c}
\hline 
Obs ID &  $T_{\rm{obs}}^{\rm \dagger}$ (ks) &  $T_{\rm{filt}}^{\rm \dagger}$ (ks) &   Date\\
\hline
0800580101 & 78.9/78.8/77.3 & 59.1/61.7/53.4 &  2017 Dec 23  \\
0800580201 & 85.6/85.6/84.0 & 50.8/54.6/29.2 &  2018 Jan 4  \\ \hline
\end{tabular} 
\end{minipage}
$^{\rm \dagger}$ The exposure times are given for EPIC MOS1, MOS2, and PN.
\label{tab:list2}
\end{table}

DF~44 was targeted by two \textit{XMM-Newton} observations for a total of 164.5~ks. Due to the large field of view of \textit{XMM-Newton}, one of the observations (ObsID: 0800580201) also includes DF~X1. However, I note that DF~X1 is at the edge of the detector, implying that the effective exposure time is significantly decreased due to vignetting effects.  I combined data from the  EPIC-PN and the two EPIC-MOS cameras to maximize the signal-to-noise ratios. The data analysis was carried out following \citet{2019ApJ...879L..12K} using the Science Analysis System software package.

The first step of the analysis was to exclude high background periods using a two step approach. First, I excluded flares using the $12-14$~keV band, which was then followed with filtering any residual flares in the $0.3-10$~keV band. To this end, a $2\sigma$ filtering was applied for both energy ranges. The effective exposure times are listed in Table \ref{tab:list2}. In addition, I excluded the out-of-time events, and those events that are  at the border of the charge-coupled devices. Throughout the analysis, only pattern zero events were utilized for all three cameras.  I identified luminous point sources and excluded them from the analysis of the UDGs. The exposure maps were created in the $0.5-8$~keV band.

\subsection{Predicted X-Ray Emission from DF~44 and DF~X1}
\label{sec:model}

The overall X-ray emission observed from galaxies originates from a multitude of sources. The most notable contributions arise from X-ray binaries and truly diffuse  gaseous emission, while other X-ray emitting components (such as accreting white dwarves or coronally active binaries) play a lesser role \citep{2004MNRAS.349..146G,2007A&A...473..783R,2008MNRAS.388...56B,2011A&A...533A..33Z,2011MNRAS.418.1901B}. 

Due to their low star formation rates, DF~44 and DF~X1 are not expected to host a substantial population of high-mass X-ray binaries  \citep{2020MNRAS.497.2759L}.  Given the low stellar mass ($M_{\rm \star} \sim 3\times10^8 \ \rm{M_{\odot}}$) of DF~44 and DF~X1 \citep{2019ApJ...880...91V}, these galaxies are also not expected to host a large number of LMXBs that were formed through the primordial channel. According to the LMXB luminosity function of \citet{2004MNRAS.349..146G}, the total expected X-ray luminosity from these LMXBs is $\sim2.4\times10^{37} \ \rm{erg \ s^{-1}}$. However, the population of LMXBs in GCs formed through dynamical interaction is expected to provide a substantial contribution given the large GC population of UDGs. Based on the luminosity function of GC-LMXBs \citep{2011A&A...533A..33Z} and the number of GCs in DF~44 and DF~X1, a luminosity of $4.3\times10^{39} \ \rm{erg \ s^{-1}}$ and $3.5\times10^{39} \ \rm{erg \ s^{-1}}$ is expected, respectively. 

Emission from the hot gas is also expected to be significant for galaxies with massive dark matter halos. Because DF~44 and DF~X1 are believed to reside in $M_{\rm vir} \sim 5\times10^{11} \ \rm{M_{\odot}}$ dark matter halos, based on the $L_{\rm X} - M_{\rm tot}$ relation \citep{2013ApJ...776..116K,2018ApJ...857...32B}, we expect the gaseous component to have an X-ray luminosity of $1.8\times 10^{39} \ \rm{erg \ s^{-1}}$. 

Thus, the total expected X-ray emission from DF~44 and DF~X1 is $6.1\times 10^{39} \ \rm{erg \ s^{-1}}$ and $5.3\times10^{39} \ \rm{erg \ s^{-1}}$, respectively.

\section{Results}
 \subsection{Images}

In Figure \ref{fig:df44}, the $0.5-8$~keV band \textit{Chandra} and \textit{XMM-Newton} X-ray images of DF~44 are presented. These images do not show a luminous X-ray glow, originating from the population of unresolved LMXBs and from truly diffuse gaseous emission, associated with the UDG. In addition, no bright point sources are associated with the galaxy.

The non-detection of luminous X-ray emission suggests that the luminosity of the X-ray emission associated with the UDGs remains below the detection threshold.

\subsection{X-Ray Point Sources}

Thanks to the sensitive \textit{Chandra} and \textit{XMM-Newton} observations, luminous X-ray point sources, such as an active galactic nucleus (AGN) or LMXBs, can be resolved in and around DF~44. To estimate the source detection sensitivity, I assume five and 20 net counts as the detection threshold for \textit{Chandra} and \textit{XMM-Newton}, respectively. Using a typical power-law  spectrum with a slope of $\Gamma =1.56$, the source detection sensitivities are $L_{\rm 0.5-8keV} = 8\times10^{38} \ \rm{erg \ s^{-1}}$ and $L_{\rm 0.5-8keV} = 4\times10^{38} \ \rm{erg \ s^{-1}}$,  for  \textit{Chandra} and \textit{XMM-Newton} respectively. Given the luminosity function of LMXBs, this allows the detection of low-luminosity AGN and bright LMXBs. 

The X-ray images of DF~44 do not reveal any luminous X-ray sources in the vicinity of the UDG (Figure \ref{fig:df44}). The search detection tools also did not detect a luminous X-ray source either in the center or in the halo of DF~44. Based on the X-ray luminosity function of GC-LMXBs, it is expected that the brightest LMXBs could be individually resolved. Given the \textit{Chandra} and \textit{XMM-Newton} source detection thresholds and the LMXB luminosity function established in  \citet{2011A&A...533A..33Z}, it is expected that $\sim0.001$ and $\sim0.005$ GC-LMXBs will be detected per GC, respectively. Since DF~44 hosts about 74 GCs, the detection of  $0.074$ and   $0.37$ GC-LMXBs is detected, implying a detection likelihood of $\sim10-30\%$. Thus, the non-detection of  X-ray sources is consistent with the source detection threshold. 

The nucleus of DF~44 does not exhibit an X-ray luminous source in its center, implying that any X-ray source is fainter than the detection threshold of $L_{\rm 0.5-8keV} < 8\times10^{38} \ \rm{erg \ s^{-1}}$. DF~44 was also observed by the Karl G. Jansky Very Large Array (VLA) in the framework of the Faint Images of the Radio Sky at Twenty-cm (FIRST) project. The $1.4$~GHz VLA FIRST image does not reveal a radio source associated with the nucleus of the galaxy with the detection threshold of $0.97$~mJy. The X-ray and radio non-detections suggest that DF~44 does not host a luminous AGN, which result is consistent with the low AGN occupation rate of UDGs  \citep{2020ApJ...898..164K}. 

Based on the fundamental plane of black hole (BH) activity \citep{2003MNRAS.345.1057M}, I derive an upper limit on the mass of an active  BH that may reside in the center of DF~44. To this end, I convert the  upper limit on the X-ray luminosity to the $2-10$~keV band and the radio luminosity to $5$~GHz. Assuming a power-law spectrum with a slope of $\Gamma=1.7$, the $2-10$~keV X-ray upper limit is $L_{\rm 2-10keV} < 5.9\times10^{38} \ \rm{erg \ s^{-1}}$. I convert the $1.4$ GHz VLA FIRST detection limit to $5$~GHz assuming a power-law spectrum, $F_{\nu} \propto \nu^{-\alpha_{\rm R}}$, where the radio spectral index is $\alpha_{\rm R} =  0.8$ \citep{2007ApJ...658..815S}. This results in a radio luminosity limit of $L_{\rm 5 GHz} < 2.2\times10^{35} \ \rm{erg \ s^{-1}}$. By utilizing the fundamental plane relation along with the X-ray and radio non-detections, I obtain an upper limit of $M_{\rm BH} < 1.2\times10^6 \ \rm{M_{\odot}}$. This limit is comparable to BH masses inferred for dwarf galaxies \citep{2020ApJ...898L...3B}. However, the non-detection of an AGN is also compatible with a scenario, in which DF~44 hosts a more massive dormant, i.e. non-accreting, BH. Indeed, the low star-formation rate of DF~44 implies that only low amounts of cold gas may be available to feed the BH, which, in turn, may result in low X-ray and radio luminosities even for a massive BH.

\subsection{Unresolved X-Ray Emission}

To qualitatively measure the X-ray emission associated with UDGs, I carry out X-ray photometry in the $0.5-8$~keV band. The source aperture is defined as a circular region with a radius of $5$~kpc (or $10.05\arcsec$). This region should encloses the bulk of the emission from the hot gaseous component and from GC-LMXBs \citep{2017ApJ...844L..11V}. The background is extracted from circular annuli with $27.5-40$~kpc (or $55\arcsec-80\arcsec$). 

The photometry confirms the empirical results based on the X-ray images: no statistically significant X-ray emission is detected either around DF~44 or DF~X1. In the absence of detections, I compute $2\sigma$ upper limits on the X-ray luminosity of these UDGs. For DF~44 the \textit{Chandra} and \textit{XMM-Newton} upper limits are $<6.4 \times 10^{38} \ \rm{erg \ s^{-1}}$ and $<2.0 \times 10^{38} \ \rm{erg \ s^{-1}}$, respectively. Although DF~X1 is not covered by \textit{Chandra} observations, the \textit{XMM-Newton} upper limit is  $<1.4 \times 10^{39} \ \rm{erg \ s^{-1}}$. These values are comparable, albeit somewhat  higher, than those obtained for the average population of UDGs in isolated and in galaxy cluster environments and exceed the X-ray luminosities predicted for a galaxy with a massive dark matter halo.

\section{Summary and Discussion}

In this work, I studied the X-ray emission arising from two archetypal UDGs, DF~44 and DF~X1. Optical observations suggest that these galaxies may reside in massive dark matter halos, similar to those found around  $L_{\star}$ galaxies. Given their low stellar mass ($M_{\rm} \sim 3\times10^8 \ \rm{M_{\odot}}$) and potentially high virial mass ($M_{\rm vir} \sim 5\times 10^{11} \ \rm{M_{\odot}}$), it was hypothesized that these galaxies are virtually ``dark'' galaxies. However, the X-ray observations presented in this Letter are inconsistent with this picture. If DF~44 and DF~X1 were residing in a massive dark matter halos, they should exhibit X-ray luminosities of  $6.1\times 10^{39} \ \rm{erg \ s^{-1}}$ and $5.3\times10^{39} \ \rm{erg \ s^{-1}}$. The observed $2\sigma$ upper limits on the X-ray luminosities are factors of $\sim30$ and $\sim4$ times lower than those expected for DF~44 and DF~X1, respectively. The X-ray faint nature of these UDGs is consistent with the low X-ray luminosities observed for nearby dwarf galaxies, such as the Large Magellanic Cloud or M\,32 \citep{2001ApJS..136...99P,2007A&A...473..783R}. 

The virial masses inferred from the present X-ray analysis and from the population of GCs is contradictory: X-ray observations suggest factor of at least $\sim5$ times lower virial mass. To understand this discrepancy, I briefly discuss the importance of metallicity in the formation efficiency of GC-LMXBs and the recent re-investigation of the GC population around DF~44. 

Galaxies hosting metal-poor GCs are less effective in forming LMXBs \citep{2003ApJ...589L..81K}. Given that DF~44 has low metallicity \citep{2018ApJ...859...37G} and most of its GCs are metal-poor, it may host factor of about three times fewer GC-LMXBs than metal-rich GCs. Taking this correction at the face value, the \textit{XMM-Newton} upper limits are still factor of $\sim10$ times lower than the X-ray luminosity expected from a galaxy with a large GC population. In reality, the required correction is significantly lower than factor of three as the GC-LMXB luminosity function of  \citet{2011A&A...533A..33Z} includes both metal-poor and metal-rich GCs. In addition, the low metallicity of DF~44 will not influence the $L_{\rm X} - M_{\rm tot}$ scaling relation, hence DF~44 and DF~X1 are still expected to host a luminous gaseous halo if they reside in massive dark matter halos.  Thus, the X-ray upper limits remain inconsistent with the presumed large GC population of these UDGs. 

The GC population of DF~44 was recently re-investigated in \citet{2020arXiv200614630S}, who suggested that DF~44 hosts $N = 19\pm5$ GC, which is only $\approx25\%$ of that measured by \citet{2017ApJ...844L..11V}. The lower number of GCs stems from a different treatment of the spatial distribution of GCs and from a different background correction technique applied by \citet{2020arXiv200614630S}. Given the $N_{\rm GC} - M_{\rm vir} $ scaling relation \citep{2020AJ....159...56B}, the population of $N = 19\pm5$ GC suggests a virial mass of $M_{\rm vir} = (9.5\pm2.5) \times10^{10} \ \rm{M_{\odot}}$. I note that this virial mass is also compatible with that obtained from the stellar velocity dispersion of DF~44 \citep{2017ApJ...844L..11V}. The low virial mass of DF~44 is comparable with the virial mass of dwarf galaxies \citep{2017MNRAS.467.2019R} and is much lower than that of $L_{\star}$ galaxies. While a similar follow-up study has not been carried out for the GC population of DF~X1, \citet{2017ApJ...844L..11V} assumed the same correction, $N_{\rm GC}  = 4 N_{\rm GC,obs}$, between the total and observed number of GCs. However, as discussed in \citet{2020arXiv200614630S}, this approach may significantly overestimate the number of GCs. 

In this work, I relied on the $N_{\rm GC} - M_{\rm vir} $ and $L_{\rm X} - M_{\rm tot}$ scaling relations to constrain the total gravitating mass of UDGs. To probe whether the conclusions are affected by the accuracy and intrinsic scatter of these relations, I briefly overview these relations. The  $N_{\rm GC} - M_{\rm vir} $ relation is extremely tight and exhibits low, $0.25$~dex, scatter \citep{2020AJ....159...56B}. This relation and its scatter is in good agreement with similar studies that connect the virial mass with either the number of GCs or the derived mass of GCs \citep[e.g.][]{2017ApJ...836...67H}. The scatter of the $L_{\rm X} - M_{\rm tot}$ relation is $0.5$~dex \citep{2013ApJ...776..116K} and its slope and normalization is in agreement with that established in \citet{2018ApJ...857...32B}. Given the X-ray upper limit on DF~44 and the intrinsic scatter of the above discussed scaling relations, DF~44 should be $\gtrsim3\sigma$ and $\gtrsim5\sigma$ outlier from the $L_{\rm X} - M_{\rm tot}$ and $N_{\rm GC} - M_{\rm vir} $ relation if it resides in a massive dark matter halo.

Overall, the lack of X-ray emission from DF~44 and DF~X1 argues that they are not ``dark'' galaxies, but they follow the stellar mass-halo mass relation established for dwarf galaxies. Therefore, it is likely that DF~44 and DF~X1 belong to the population of dwarf galaxies. This result is consistent with that established for the bulk of UDGs residing in isolated and cluster environments. Thus, it is unlikely that even a small subset of UDGs are failed $L_{\star}$ galaxies and suggests that UDGs comprise a single population.

\bigskip

\begin{small}
\noindent
\textit{Acknowledgements.}
I thank the referee for the constructive report and Andra Stroe for helpful discussions on the radio observations of DF~44. The scientific results reported in this article are based to a significant degree on data obtained from the Chandra Data Archive and software provided by the Chandra X-ray Center (CXC) in the application package CIAO. This work uses observations obtained with \textit{XMM-Newton}, an ESA science mission with instruments and contributions directly funded by ESA Member States and NASA.  The National Radio Astronomy Observatory is a facility of the National Science Foundation operated under cooperative agreement by Associated Universities, Inc. \'A.B.\ acknowledges support from the Smithsonian Institution.

\end{small}

\bibliographystyle{apj}
\bibliography{bib2} 

\end{document}